# Room temperature ballistic conduction in carbon nanotubes

Philippe Poncharal[†], Claire Berger[††], Yan Yi, ZL Wang, Walt A de Heer*

School of Physics – Georgia Institute of Technology- Atlanta GA 30332

**Abstract**

Multiwalled carbon nanotubes are shown to be ballistic conductors at room temperature, with mean free paths of the order of tens of microns. The measurements are performed both in air and in high vacuum in the transmission electron microscope on nanotubes that protrude from unprocessed arc-produced nanotube containing fibers which contact with a liquid metal surface. These experiments follow and extend the original experiments by Frank et al (Science, **280** 1744 1998) that demonstrated for the first time the large current carrying capability, very low intrinsic resistivities, and evidence for quantized conductance. This indicated 1D transport, that only the surface layer contributes to the transport, and ballistic conduction at room temperature. Here we follow up on the original experiment including in-situ electron microscopy experiments and a detailed analysis of the length dependence of the resistance. The per unit length resistance $\rho < 100$ $\Omega/\mu m$, indicating free paths $l > 65$ $\mu m$, unambiguously demonstrate ballistic conduction at room temperature up to macroscopic distances. The nanotube-metal contact resistances are in the range 0.1-1 k$\Omega\mu m$. Contact scattering can explain why the measured conductances are about half the expected theoretical value of 2 $G_0$. For V>0.1V the conductance rises linearly (dG/dV~0.3 $G_0$/V) reflecting the linear increase in the density-of-states in a metallic nanotube above the energy gap. Increased resistances ($\rho$ =2-10 k$\Omega/\mu m$) and anomalous I-V dependences result from impurities and surfactants on the tubes. Evidence is presented that ballistic transport occurs in undoped and undamaged tubes for which the top layer is metallic and the next layer is semiconducting. The diffusive properties of lithographically contacted multiwalled nanotubes most likely result from purification and other processing steps that damage and dope the nanotubes thereby making them structurally and electronically different than the pristine nanotubes investigated here.

**Introduction.**

It can hardly be argued that the most fundamental electronic transport property is electrical conductivity, and that the discovery of novel conductivity phenomena should be regarded as extremely important. In 1998 Frank et al.[1] provided strong evidence for room temperature ballistic conduction on micron length scales in pristine freely suspended carbon nanotubes. They concluded that the multiwalled carbon nanotubes (MWNTs) are one-dimensional conductors, that electronic transport occurs on the outer layer and that current densities greater than those observed in any material (excluding superconductors) could be attained. These multiple observations were all novel. Ballistic transport at room temperature over microns distances was unknown for any system and had not been observed in nanotubes of any kind, not even at low temperatures. The evidence came from a deceptively simple experiment where MWNTs were brought into contact with a liquid metal and the resistance was measured as a function of the depth that the nanotube was plunged into the liquid metal[1,2,3]. The conductance appeared to be quantized with conductance values remarkably close to the quantum of conductance $G_0 = 2e^2/h \approx 1/13 k\Omega$ and virtually independent of the depth that it was submerged. All of these effects, including the diameter independence of the effect, supported room temperature ballistic conduction over microns distance[1]. The observation was all the more surprising since it was at odds with other experiments at the time, moreover none of the observed effects had been predicted for MWNTs. In fact, prior experiments showed that MWNTs were diffusive, 3 or 2 dimensional conductors[4,5,6,7] exhibiting diverse transport properties. There was no indication that the transport was confined to the outer layer, that MWNTs could sustain large currents, or that they were one-dimensional conductors or room temperature ballistic conductors. However, Tans et al had reported one-dimensional coherent transport in single walled nanotubes (SWNTs) on the 200 nm length scale at cryogenic temperatures[8].

Subsequently, room temperature ballistic conduction has been verified for SWNTs on the 200 nanometer scale[9] and at low temperatures on the micron scale[10,11]. In contrast, room temperature ballistic conduction in MWNTs has been negated by several of investigators, in experiments involving lithographically contacted nanotubes. Instead, multishell conduction[12], low-temperature quasi-ballistic conduction (with mean free paths of the order of 150 nm)[13], and diffusive conduction[14] has been reported for MWNTs[15]. In a recent development, quantum dot properties were observed in a MWNT at low temperatures[16] quite (similar to those observed by Tans et al in SWNTs[8]) indicating long coherence lengths. However it was concluded that the coherently transporting layer was not on the top but a submerged layer and that the top layer was diffusive due to doping by atmospheric water. Hence those MWNTs exhibit both diffusive and ballistic properties simultaneously. Although this explanation may reconcile some experimental observations involving processed lithographically contacted nanotubes, it still does not explain why our experiments show ballistic properties in the top layer and the very long mean free paths.

The explanation must be sought in the processing of the nanotubes. Whereas Frank et al *specifically avoided all processing in order to avoid contamination and damage*[1,3,2], MWNTs continue to be processed before they are measured in most other experiments. In particular, ultrasonification and surfactant stabilization[17,18,19,20] are used almost universally prior to the deposition and application of the electrodes. Others involve thermal annealing at T~450 C to burn out carbonaceous particles [21,22]. Because the transport is known to be primarily on the surface[1,13,23] it is hardly surprising that these treatments, which directly affect the surface and which are known to be damaging[24,25,26,27,28], indeed significantly affect the transport properties, and hence provide a simple explanation for the discrepancies in the experiments. We also emphasize that the MWNTs discussed here are produced in pure carbon arcs [29]: catalytically produced MWNTs are highly defective [30] and do not have the ballistic transport properties discussed here [31].

These discrepancies are not subtle but vast. Whereas the resistances per unit length of lithographically contacted nanotubes are found to be $\rho = 10$ k$\Omega/\mu m$[14], we show here that $\rho < 100$ $\Omega/\mu m$ for our freely suspended MWNTs, which implies mean free paths of the order of 100 $\mu m$ (rather than at most a few hundred nm).

Here we present in detail the properties of freely suspended MWNTs, expanding on the methods developed in the original experiment[1,2]. In particular, the length dependence of the conductance of the nanotubes is carefully analyzed and the contributions of the contact and from scattering along the nanotube are identified. These measurements reveal that the nanotube contact resistances are large and that scattering in the nanotube is so small that it approaches the uncertainty of the measurement (which is in the range of tens of Ohms per micron). We demonstrate that the conductance increases linearly with increasing voltage at high bias and that this effect is directly related to the density of states (or more aptly, due to the opening of higher conducting channels, which however have small transmission coefficients). Currents of the order of mA are routinely achieved. We further present the results of extensive measurement in-situ electron microscopy experiments, which show the effects of impurities and damage on the conductivity. The effects of surfactants on the conductance are also shown. The final picture is relatively simple: Undamaged multiwalled carbon nanotubes with a metallic outermost layer are room temperature ballistic conductors over distances which may exceed 100 µm in ambient conditions. Only the outer layer participates in the transport. The higher subbands contribute minimally to the conductance of (long) nanotubes, even at high bias and with significant doping. The reduction of the conductance from 2 $G_0$ to 1 $G_0$ is probably due to scattering at the second contact.

**Experiment**

The basic experiment has been described in Ref.[1] [2] Nanotubes are produced using the pure carbon arc method [29]. Power to the arc is supplied from a full-wave rectified AC supply (20 V, 80 A); the arc is struck in the 1mm gap between a 7 mm diameter graphite anode and a 5 cm diameter graphite cathode in a 500 Torr He atmosphere. The MWNTs (with diameter D= 5-25 nm and length L=1-10 µm) are found on the anode in a soft sooty deposit inside a hard carbonaceous shell. The soot is composed of loosely packed fibers that are approximately aligned with the arc. The fibers consist of compacted MWNTs (~80%) and other graphitic objects (amorphous flakes and polyhedral particles which cover the nanotubes). The fibers are typically 1 mm long and 0.1 mm in diameter[29]. Microscopic investigation shows that nanotubes protrude from the fiber.[1] [32]

A fiber is carefully separated from the deposit and attached to a conducting wire using silver epoxy and attached either to the modified probe of an scanning probe microscope (SPM, Park Instruments Autoprobe CP)[1] or to a manipulator in the transmission electron microscope (TEM)[32] (see Fig. 1). Using the SPM, the fiber is lowered and contacted to a liquid metal surface (typically Hg, Ga and various low melting point metals have been also used[1]; experiments with Ga are performed under high purity silicon oil). A voltage V~ 100 mV is applied to the tip and the current I is measured using a fast transient digitizing oscilloscope (LeCroy 574AM) that also records the position of the nanotube with respect to the liquid metal surface. The data are recorded at rates corresponding to 10-50 channels/nm (the record of a single trace consists of 100.000 data points). Contact of the nanotube with the liquid metal surface results in a jump in the conductance. The conductance G(x)=I/V is measured as a function distance x that the fiber is lowered into the liquid metal (see Fig. 2). The successive steps in a trace result from several tubes that successively come into contact with the mercury. The experiment is repeated at a typical repetition of 1-10 Hz for typically several hundreds to several thousands of cycles where the tip is raised and lowered by in the range Δx=1-10 µm. Initially the steps in G(x) are poorly defined (Fig. 2a,b) and a dark deposit is found to appear on the metal surface, which can be observed with the SPM alignment microscope. This deposit comes from material from the fiber (as verified in in-situ TEM experiments). The fiber is then displaced to a fresh area of the metal. A stable pattern of steps is established after some time (Fig. 2c), which typically reproduces for at least several hundred cycles. While the plateau lengths may vary somewhat from one cycle to the next, the values of the conductances at the steps are stable within about 5% (see Ref [1]). In air, oxide layers build up on the Hg surface after about 1 hour, where after the surface is cleaned. Data is automatically collected in sequences of 50 or 100 traces.

The effect is robust and produces results related to those discussed here in most of the cases. Occasionally the experiment fails to produce steps and the conductance jumps immediately to full contact (10-100 Ω). TEM examination of some of these tips showed that there were no tubes extending from the fiber. Frequently the nanotubes at tips of virgin fibers are coated with a thin layer of amorphous carbon and amorphous carbon balls (which are currently under investigation), which have been correlated with anomalous nanotube conductances. For this reason, the tips of the fibers are carefully removed to expose the nanotubes inside.

Two point current-voltage (I-V) measurements are made by sweeping the voltage and recording the current, either continuously (using a fast high-resolution digitizer) or point by point.

TEM measurements are conducted similarly where the nanotube fiber is connected to a manipulator so that its position with respect to a liquid metal coated copper wire can be adjusted, however in this case the manipulation is done manually. The TEM measurements are primarily performed to characterize the condition of the fibers and to verify the processes observed in the in-air experiments. Because it is very difficult to align to the electron beam with respect to the fiber-metal contact in order to observe the contact point only a limited number of measurements have been made in this mode. The measurements are in agreement with the more extensive measurements in air.

**Results**

The evolution of the steps with cycling time is shown in Fig. 2. Defined and reproducible conductance steps usually evolve only after repeated dipping into the liquid metal. Initially the steps are poorly defined with large slopes. The slope of the first step in Fig 2a corresponds to –dR/dx =36 kΩ/µm; the slope of the plateau in Fig. 2b corresponds to –dR/dx =4 kΩ/µm. The ultimate conductance plateaus are very flat with some rounding at the steps. The typical ultimate conductance values of the first plateau ranges from $G_{pl}$=0.5-1 $G_0$. Sometimes even lower values are seen (see below), however initial plateaus with $G_{pl}$ substantially greater than 1$G_0$ are not observed. A typical conductance trace consists of several upward conductance steps when the fiber is pushed down. The sequence is reversed when the fiber is retracted. Typically $G_{pl}$ varies slightly from trace to trace (by a few percent).

Electron microscopy studies reveal that as the nanotube is pulled away from the surface, just prior to breaking contact, a cone shaped meniscus is drawn from the Hg (Fig. 1a inset) This causes an offset of the position the step going into the Hg compared with coming out (Fig. 1b). This effect is due to non-adhesive wetting (a simple experiment with a glass rod touching a mercury surface demonstrates this effect). Neither mercury nor gallium wet nanotubes [33][34][35]. However due to the effect mentioned above, we only analyze conductance traces going into the Hg and not as the tubes are withdrawn.

Poorly defined steps correlate with the degree of contamination on the nanotubes: nanotubes that have not been in contact with Hg tend to be covered with graphitic particles as can be seen in the electron microscopy images (Fig. 2d). The dipping process initially causes some changes in the morphology of the nanotubes protruding from the fiber. In particular some tubes move from their original position. Occasionally large fragments are transferred from the fiber to the metal (as observed in the TEM). The evolution of the steps and the TEM images (Fig. 2a-e) suggest that the dipping process not only cleans the graphite particles from the tubes but also insures that only those nanotubes that are well anchored remain in place. The former process causes the plateaus to become flatter and less noisy whereas the latter process raises the plateau value, when better contact of the nanotube with the fiber is established.

A typical conductance trace is shown in Fig. 3a, and consists of a rapid rise at x=0, followed by a rounded step with a flat plateau near $G_{pl}$=1 $G_0$. A detailed analysis is given below. We have observed that the nanotubes that protrude from specific fibers often produced flat conductance plateaus with significantly lower conductance values (about 0.3-0.5 $G_0$). These have been attributed to poor contacts with the fiber, since these plateaus are prone to jump to larger values and ultimately to stabilize.

The effect of surfactants and solvents has been investigated. Nanotube fibers were dipped in an aqueous solution of sodium dodecyl sulfate (SDS) and dried (SDS is a surfactant that is often used to suspend nanotubes[17]). Prior to this treatment these fibers produced the typical flat plateau structures. Surfactants affect the conductance properties of the nanotubes; figure 4a shows a typical conductance trace (one of 100 recorded in this series). The steps of the surfactant coated tubes have reduced conductances and the resistance decreases linearly with increasing x (figure 4b) indicating a resistive nanotube, as discussed below.

Conductance G=I/V versus voltage measurements of clean nanotubes show a typically symmetric pattern (see Fig. 5). At low bias the conductance is constant, up to about 100 mV, whereafter it rises with a constant slope; typically dG/dV =0.3-0.5 $G_0$/V. In contrast to SWNTs[36], saturation of the current (or conductance) is not seen[2], at least up to |V|= 4 Volts where I=615μA (see Fig. 5a). Measurements made in the electron microscope confirm the in air results. For example, a 15 nm diameter, 0.5 μm long nanotube measured in-situ, shows a linear rise of about 0.5 $G_0$/V for V>0.2V (Fig. 5a, inset).

Fig. 5b shows G(V) measurements of a nanotube submerged to various depths in the liquid metal, which do not to show significant changes from one depth to the next.

Nanotubes coated with a surfactant also have anomalous G(V) properties. An example is shown in Fig. 6. The conductance rises but not linearly nor is it symmetric with respect to V=0. This behavior is representative of the modification observed with surfactants which show various degrees of asymmetries and shifts compared with the pure case.

In situ TEM investigations show that not all nanotubes conduct. Nanotubes that are clearly in contact with the liquid metal may exhibit resistances above our measurement limit of 1 MΩ. Moreover, these nanotubes (~1 μm from contact to contact) typically can withstand voltages exceeding 5V. This indicates that these tubes are robust insulators; apparently tunneling into deeper metallic layers is inhibited.

Large applied voltages destroy conducting nanotubes. Fig. 7e-f shows the result of passing I> 1 mA current through the nanotube. The surface of the nanotube is disrupted and has been damaged by the current along its entire length. From this low-resolution image it is estimated that less than 3 layers are effected. These properties are typical and others have obtained similar results[23][37].

High voltages applied to defective tubes cause them to break at the defects as shown in the TEM micrographs in Fig. 7, whereas non-defective tubes tend to break at or near a contact point. We specifically have not observed undamaged nanotubes which broke in the middle as observed in Ref[38].

**Analysis of conductance curves**

Below we present a detailed analysis of the conductance traces. The plateau curvature is determined by the increase in the total conductance of the system as the contact area with the Hg becomes larger and at the same time as the distance along the nanotube from contact to contact becomes shorter. We only analyze the first steps (that is the step following the first significant rise from G=0, which corresponds to one nanotube in contact with the Hg), and not the subsequent ones.

The conductance properties of nanotubes are reflected in the conductance traces G(x) where x is the displacement of the SPM, x=0 corresponds to the point where contact is made (i.e. where the conductance steps up from 0). We concentrate primarily on longer plateaus to accurately quantify the resistance per unit length of the nanotube. We provide two examples in detail, one of a typical nanotube with a plateau value near 1 $G_0$, and an other with a significantly reduced plateau value. Note that these are representative results of many measurements of these plateaus. These experiments and their analysis have been carried out over the past four years and the results we present here are clear examples of typical behavior.

Figure 3a shows a conductance trace with a step and a plateau value of $G_p$ =0.9 $G_0$. This trace is typical of the 50 recorded traces of this plateau. This plateau is long and extends for 2.5 μm and has the characteristic rounded shape close to x=0.

Fig 3b shows the same plateau, but now represented in terms of resistance R(x)=1/G(x). The length of the exposed nanotube outside the fiber is L. The nanotube resistance per unit length is ρ. The combined contact resistances are $R_C$ so by Ohms law, the total resistance is

R(x)= $R_C$ +(L-x) ρ         (1)

dR(x)/dx=d$R_C$(x)/dx- ρ         (2)

Since R(x) is not at all represented by a straight line indicates that the contact resistance also depends on x. Continuing classically, the metal contact resistance varies inversely proportional to the contact area, hence $R_{C1}$=$R^*_{NT-M}$/x, while the resistance of the contact to the fiber is constant: $R_{C2}$=$R_{NT-F}$. In total, the classical nanotube resistance is given by

R(x)= $R_{NT-F}$ +(L-x) ρ +$R^*_{NT-M}$ /x         (3)

Because dR(x)/dx=-($R^*_{NT-M}$ /$x^2$+ ρ ), we can immediately establish an upper bound for ρ by measuring the dR(x)/dx for large x. From Eq. 3 it is clear that:

ρ < -dR(x)/dx         (4)

For this trace we find the upper bound (at x=2.5 μm): ρ <48 Ω/μm (see Fig. 3b).

Hence, from this elementary analysis of this trace we find that the contribution to the total resistance per micron length is at most 50 Ω which is a factor of 260 less than 13 kΩ (~1/$G_0$). The significance of this is presented below.

Figure 3b also shows results from Ref. [13] ( ρ=4kΩ/μm) and from Ref.[14], (ρ =10kΩ/μm) which are much larger than the upper limit found here. To refine this value, we must evaluate the contact term R*.

When ρ is small, the plateau resistance (in this classical picture, the quantum case is treated below) is approximately given by $R_{pl}$=$R_{NT-F}$, Eq. 3.

Plotting the resistance R with respect to 1/x (while appropriately accounting for the conductance step at x=0 [39]) clearly demonstrates the contact resistance effect (Fig. 3c). The result is a straight line which intercepts 1/x=0 at $R_{plateau}$ =14.1 kΩ and which has a slope R*= 270 Ω μm. This unambiguously demonstrates that the shape of the conductance step is dominated by the contact resistance and not by the intrinsic nanotube resistance ρ. The smooth line running through the experimental data in Fig. 3b represents the result of an unconstrained fit to Eq. 3 (including the $R_{tip}$, Ref. 39). For this trace ρ =14 Ω/μm. Hence, only a small fraction of the slope at x=2.5 μm can be due to the intrinsic resistance of the tube. Furthermore, the contact resistance is found to be R* =256 Ω.μm from this fit (i.e. close to the value found from the slope in Fig. 3c ).

The above procedure was incorporated in an automated fitting routine and applied to analyze the 50 measurements of this step. The slope of R at x=2.5μm is –dR/dx=87±52 Ω/μm. The unconstrained fit (which allows negative values) gives a distribution of measured values with ρ=31±61 Ω/μm. These values are typical for MWNTs investigated in this study.

Furthermore, the contact resistance found from the fit of the 50 measurements is R*=167±55Ωμm. Note that others find comparable contact resistances. In particular, Schonenberger et al [13] find 3.8 kΩ average contact resistances for 100-200 nm wide MWNTs which corresponds to a resistance per unit length: R*= 380-760 Ω μm.

Another series of 70 measurements of a 2 μm plateau with a particularly low plateau conductance (~0.5 $G_0$) similarly analyzed is presented next (Fig 4c). From the distribution of the measurement values of this plateau we find that ρ=40±45 Ω/μm and R*=1100±130 Ωμm. From the TEM we know that the nanotubes typically protrude at few μm from the fiber, so then the maximum contribution to the resistance due to the nanotube is of the order of a few hundred Ohms. Hence, this analysis shows that the reduced plateau value (i.e $G_{pl}$ ~0.5 $G_0$ rather than ~1 $G_0$) is not due to the nanotube resistance, but rather due to a larger than normal contact resistance at the nanotube-fiber contact, which is discussed in detail below.

The important message to be gained from the above is that the intrinsic MWNT resistances are very low; in fact they are orders of magnitude lower than those reported by others (both MWNTs and SWNTs). We stress that because the contribution from the contact and that of the nanotube both act to increase the conductance with increasing x, therefore the contact contribution cannot possibly compensate the resistive contribution of the nanotube. Furthermore, since the two contributions have different functional dependences on x, they can be isolated as was done in the above analysis.

**Ballistic transport in carbon nanotubes.**

Depending on the helicity (n,m), single walled carbon nanotubes are either metallic (for n=m), narrow band semiconductors (when n-m is a multiple of 3), or semiconductors[40]. Theoretically the band-gap for semiconducting nanotubes is of the order of $\Delta E_{sc}=2\gamma_0 a/D$ where $\gamma_0$~3 eV is the energy overlap integral used in tight binding calculations for graphite and nanotubes[41,42], a=0.14 nm and D is the diameter of the tube in nm. For metallic undoped tubes, two 1D subbands with a linear dispersion cross exactly at the Fermi level. These are the metallic subbands which give the tube its metallic character. Systems of unoccupied and occupied levels are symmetrically positioned above and below the Fermi level with a structure that resembles that of the semiconducting nanotubes (they do not cross the Fermi level, and hence we refer to them as the semiconducting subbands). The gap between the system of unoccupied and occupied levels in metallic nanotubes is three times as large as for the semiconducting tubes[43,44]; $\Delta E_{metal}=6\gamma_0 a/D$[45,46]. For example, for a 15 nm diameter tube, $\Delta E_{metal}$= 0.17 eV. These gap sizes have been verified experimentally by Venema et al.[42]. Note that $\Delta E_{metal} \gg kT$ for T=300 K for the typical MWNT diameters (D=5-25 nm). Hence at room temperature and for bias voltages V< $\Delta E_{metal}$ only the metallic subbands are expected to contribute to the transport.

Figure 8 shows the band structure (calculated in the tight-binding approximation which adequately describes the basic structure[45,46]) and the density of states of a (n,m)=(100,100) nanotube. This is a conducting tube with a diameter D= 13.6 nm (typical for the nanotubes in the present study) and $\Delta E_{metal}$=0.18 eV. The density of states is shown in Fig. 9. The properties of this tube are representative of all conducting nanotubes of this diameter. Note the van Hove singularities, which are largely washed out at room temperature. Also shown is the conductance as a function of bias voltage according to the Landauer equation, assuming a transmission coefficient T=1 (see below for details).

The scattering properties of the metallic subbands (the two subbands that cross the Fermi level) and the semiconducting subbands (those bands which do not cross $E_F$) of metallic nanotubes are found to be very different. For the former back scattering is forbidden due to the fact they are essentially of pure π (bonding) and π* (anti-bonding) character, in contrast to the semiconducting subbands which are of mixed character and consequently they can back-scatter[47,48]. Hence, even if the states above the gap become populated (thermally, by doping, or by large bias voltages) it should be expected that (for long nanotubes) the two conducting subbands provide the primary contribution to the current. *Hence, it is theoretically expected that the scattering in the metallic subbands of metallic nanotubes is much smaller than in the (doped or thermally populated) semiconducting subbands of the same nanotubes.*

Indeed, transport properties of metallic SWNTs and those of doped semiconducting ones have been measured[48] and the mean free paths of the latter have been found to be much shorter than those of the metallic SWNTs confirming the predicted[47] unique low scattering properties of the metallic subbands.

The band structure of SWNTs (both metallic and semiconducting) has been experimentally verified[44,43]. In their tunneling experiments, Schonenberger et al[13] have shown that the electronic density of states of MWNTs corresponds to the theoretical predictions. It is similar in structure to a SWNT however with the expected reduced gap size due to the larger diameters. Bachtold et al.[49] also demonstrated that only the top layer participates to the transport (at least at low temperatures).

According to the Landauer equation[50,51], in absence of scattering and with perfect contacts the conductance of a system with N conducting subbands or channels is $NG_0$. This ideal is not met in real systems. Accordingly

$G=G_0\Sigma T_I$ (5)

Where the sum is over the transmission coefficients (0≤$T_I$≤1) of the conducting channels. For an ideal nanotube with ideal contacts the transmission coefficient for both channels equals unity so that G=2 $G_0$. In the non-ideal case the transmission is reduced, due to back-scattering in the tube and imperfect contacts. When the scattering length in the nanotube is much greater than the intercontact distance, then the conductivity becomes independent of the length and the nanotube is considered to be a ballistic conductor[50,51].

The mean free path in this context refers to the momentum scattering length $l_m$, which includes any process that alters the electronic momentum and hence affects the resistance.

The intrinsic resistance of the nanotube (due to scattering) is related to the transmission probability using the four-terminal Landauer formula[14 51 52] assuming 2 conducting channels:

$$R_{intr}=(h/4e^2)(1-T)/T \qquad (6)$$

where T is the transmission coefficient for electrons along the length of the nanotube. Following Bachtold et al[14], ballistic transport is unambiguously demonstrated when T>1/2 because then the majority of electrons traverse the nanotube without scattering. Because R=L ρ this criterion is satisfied up to distances $L_{max}=(h/4e^2)/\rho$.

Explicitly, the transmission coefficient of a 1D wire with scattering centers (ignoring quantum interference effects) is given by (see i.e. Datta,[51] p.62)

$$T_S=(1+L/L_0)^{-1} \qquad (7)$$

where $L_0$ is of the order of $l_m$. ( Note that in the diffusive limit, including multiple reflections, for $L_0/L \ll 1$, $G=NG_0L_0/L$ where N is the number of channels, which is consistent with Ohms law [51, 14, 48]). Hence the total resistance for the nanotube, assuming two conducting channels and including the contact conductance $G_C$ is given by (see i.e. Datta [51] p.62)

$$R(L)=G^{-1}=G_C^{-1}+G_S^{-1} = (2G_0)^{-1}(T_C^{-1}+L/L_0) \qquad (8)$$

Here $T_C$ is the transmission coefficient through both contacts. This expression relates the mean free path to the tube resistance, yielding a linear dependence of the resistance on the nanotube length as in the classical case (Eq. 1). Hence, the term linear in L can be directly compared with the experimentally determined value (Eq.3):

$$(2G_0)^{-1}/L_0 = \rho \qquad (9)$$

(The mean free path can also be found from the Einstein relation[53], as shown in Ref.[3]). As explained above, the slope in the conductance trace provides an upper limit for ρ. Consequently, for the plateau of Fig, 3, $\rho < 50$ Ω/μm and hence $L_{max}$=130 μm. Nanotubes shorter than this are room temperature ballistic conductors over their entire length. Correcting for the contact resistance (as shown above) yields ρ =31±61 Ω /μm, which implies

$L_{max}$~200 μm.  (corrected for contact resistance)

This implies that MWNTs are ballistic conductors at room temperature for lengths up to at least a fraction of a mm.

The results found here are typical for the nanotubes studied. Hence MWNTs are not only unambiguously room temperature ballistic conductors, but over unprecedented distances. The results cannot coherently be explained in term of multiple conducting channels (with reduced transmission coefficients). First, in order to have diffusive behavior with ρ =100 Ω/μm with a mean free path of the order of $l_m$ = 0.2 μm (which is the quasi ballistic scattering length quoted in Ref. [13]) would require by Landauer Buttiger theory[51 54].

N=( $l_m \rho G_0)^{-1}$=650 Channels.

In contrast, the number of participating channels is experimentally[13 55] found to be of order unity (as expected theoretically as well) even for deliberately heavily doped samples[56], so that explanations of the low resistances that involve many channels with small mean free paths are unfounded and not based on the well understood and accepted nanotube properties.

Second, the measured two point conductances are always near 1 $G_0$. There is no physical reason the contact resistances of spuriously doped nanotubes with a large variety of diameters would exhibit such an effect.

**Scattering at Contacts**

We find that G≤1$G_0$ and that values near G=0.9±0.1 $G_0$ are the most common. We have conducted these experiments for several years with several investigators. We have recorded many cases for which the plateau conductances $G_{plateau}$<1$G_0$, and examples are given here. In particular, G~0.5 $G_0$ are observed relatively frequently, although these plateaus often (but not always) abruptly progress to plateaus near 1$G_0$. In contrast, we have not observed initial conductance steps that are significantly greater that 1$G_0$. Furthermore, the plateaus are invariably flat (not sloped) with a rounded step. Hence, even allowing for a distribution in plateau values, the cutoff at 1 $G_0$ appears to indicate, as originally claimed[1], that only one quantum of conductance is involved rather than two. Possible explanations for the 'missing' quantum of conductance were pointed out in that work[1 2], and subsequently by others[57 58 59]. Several explanations addressed the properties of the metal-nanotube contact. Experimentally however, high transmission nanotube to metal contact have been demonstrated (see i.e. Ref.[10]). Below we give an explanation in terms of reflections from the nanotube-fiber contact.

To proceed, we first develop a semi-classical model for the contacts. This development, presented in the Appendix, follows that originally proposed in Ref[60 61] which is based on the Landauer- Buttiger theory[51, 54] and related to the Datta's semi-classical discussions[51]. The methods were developed to explain fractional conductances observed in gold nanowires[60] and carbon nanotube networks[60][61]. The point of the model is to find expressions for the transmission coefficients in the Landauer equation (Eq.5, see Appendix). The model assumes that the elastic scattering of an electron at interfaces and scattering centers is isotropic. Hence an incoming electron scatters with equal probability into each of the outgoing channels (similar to the isotropy condition, cf Beenakker[62]). Quantum interference effects are ignored, but multiple reflections are considered. Accordingly, the total resistance of a nanotube of length (L-x) with two conducting channels, contacted on one end to a metal contact of length x and to a non-reflecting contact at the other end (see Appendix for details), is

$R(x) = (2G_0)^{-1}((C_1 x)^{-1} + (L-x)/L_0 + 1)$ (10)

This resembles the classical Ohmic expression (Eq. 3) although it does not assume diffusive transport but rather it relies on transmission and reflection of electrons at the interfaces of the various elements. The first term represents the nanotube-metal contact; $C_1$ is an empirical constant that can be estimated from the Sharvin equation[63]

$C_1 \sim \pi r^2 / \lambda_F^2$ (11)

where r is the nanotube radius and $\lambda_F$ is the Fermi wavelength in the nanotube. With $\lambda_F \sim 40$ nm for graphite [41], and 5nm<r<10nm then the conductance of the metal-nanotube contact is $2G_0 C_1$ and 30 μm$^{-1}$ <($C_1$)<60μm$^{-1}$. The experimental values, found from the previous analysis range from 10-35 μm$^{-1}$, hence in surprisingly good agreement with this very simple estimate. The second term in Eq. 10 is due to scattering along the nanotube with a mean free path $L_0$ discussed above, the third term represents the quantization of conductance in 1D systems.

The nanotube-metal conductances found here are in line with the contact transmission coefficients calculated by Anantram et al[64] for SWNTs. In that treatment of various types of nanotubes with metals it is shown that the transmission coefficient increases linearly with contact area, hence in agreement with the semi-classical model used here.

The nanotube-fiber contact is more complex. As discussed in detail in the Appendix, it consists of a series of nanotube-nanotube contacts. In the model, an electron scatters isotropically at the junctions between nanotubes. Considering an infinite series of such junctions (as an approximation to the real nanotube-fiber contact) then the transmission probability from the nanotube to the fiber $T_{NT-F} \sim 0.7$. On the other hand, crossed nanotubes have been studied and the transconductance from metallic to metallic singlewalled nanotubes have been determined[65]. The probability that an electron on one tube tunnels from to the next is found to be about T=0.06 [65]. Using this value, we find for an array of these junctions that $T_{NT-F} = 0.3$ (see Appendix). Hence $0.3 < T_{NT-F} < 0.7$.

If we assume that two channels contribute to the transport in the nanotubes, then from the empirical values 0.5 $G_0 < G < G_0$, we conclude that T is a distribution with $0.25 < T_{NT-F} < 0.5$ which peaks at $T_{NT-F} \sim 0.5$. This may explain the origin of the missing quantum in terms of the transmission coefficient into the fiber.

**Scattering from defects and contaminants**

Scattering on the nanotubes, from static scattering sites (defects and surface contaminants), increase the resistance. As shown by Chico et al[66] a defect in an (n,n) nanotube reduces the conductance: a vacancy on a 10 nm diameter tube reduces the conductance by about $\Delta G = 0.15 G_0$ (for a 1.4 nm diameter SWNT the reduction is about 1 $G_0$). Consequently, if a nanotube with a defect is contacted with a liquid metal electrode, then the conductance should make an upward step of $\Delta G$ when the defect becomes submerged in the liquid metal (thereby shorting out its effect). These relatively large steps are readily visible in conductance traces of contaminated tubes (see Fig. 2a,b) but they are seldom seen on clean tubes. More specifically, since the plateaus of conditioned tubes are smooth indicates that they are essentially free of point defects over extended lengths (order of μm).

The relatively high frequency with which $G \sim 0.5 G_0$ plateaus are observed (cf Ref. [1] Fig. 2) deserves special note and in particular that these plateaus often evolve to stable plateaus with $G \sim 1 G_0$ during the execution of the experiment. Conductance jumps of a factor of about 2 have been observed in the TEM and they were correlated with significant changes in the contact to the fiber. In particular, 'pseudo-contacts' will reduce the transmission by a factor of two (cf [51], [9]). Hence it is likely that these reduced plateaus are due to pseudo-contacts.

Surfactants dramatically affect the transport behavior. Figure 4 shows a conductance step and its associated resistance step. Note the absence of a flat plateau. Rather the resistance changes uniformly with x and with a slope that corresponds to ρ =2.2 kΩ/μm. (The metal-nanotube contact resistance is 500 Ωμm for this step). The ρ value is at least an order of magnitude greater than observed for clean tubes. Contrary to clean tubes, the resistance is not linear with 1/x (inset of figure 4b), which indicates that the shape of the conductance is not determined only by the contact conductance. Note also that (as for clean MWNTs) the plateau is smooth, and that there is no evidence for abrupt steps that would result from strong scattering centers (as for the tubes contaminated with particles). These results demonstrate that surfactants greatly increase the resistance of the nanotube. The current-voltage characteristics are also strongly affected as discussed below.

Currents typically greater than 1 mA destroy the tubes[1] as shown in Fig. 7. Defect free nanotubes tend to shed their outer layer or layers over their entire length (Fig. 7e-f). The contact is disrupted at the liquid metal-nanotube contact. This observation (see also Ref. [23]) confirms that only the outer layer or layers participate to the transport even at high current densities. Transport in the outer layers has in fact also been concluded in by others [23][13]. The number of layers involved at higher current densities (high bias voltages) is not known but it is very likely that only the top conducting layers are involved. Because on average only one in three layers are conducting, it is expected that only one in three conducting nanotubes have two conducting top layers and one in nine have three conducting top layers and so forth. Hence in a majority of the cases, it is likely that only one layer is involved, even for high currents.

From in-situ microscopy experiments[32] (see Fig. 7) we observed that **(1)** defect free nanotubes tend to break at the contact point with the liquid metal, rather than at the nanotube-fiber contact, or in the middle of the nanotube, which would be the hottest point if it were a diffusive and dissipative conductor (as in Ref.[38]). **(2)** tubes that are coated with particles tend to break near the locations of these particles; **(3)** kinked nanotubes break at the kink. These experiments are consistent with the conclusion that dissipation occurs at defects and at contaminants.

**Conductance versus voltage**

The conductance versus voltage G(V) properties of MWNTs (Fig. 5) are summarized as follows: G(V) is essentially constant up to V~100 mV, where after is rises linearly with a slope which is typically dG/dV= 0.3 –0.5 $G_0$/V. The slope is constant up to at least V= 4V (I=0.56 mA). The curves are symmetric about the V=0 axis with a slight offset (typically less than 10 mV). The G(V) appears to be only weakly dependent on x (see Fig.5b and inset). In-situ TEM experiments also show the linear conductance increase (Fig. 5a). There is no evidence for saturation of the conductance. In particular, the current saturation effect observed in SWNTs[36], which would result in a 1/V decrease in the conductance, is not observed. For SWNTs the saturation affect is attributed to back-scattering from longitudinal phonons however apparently this does not occur in freely suspended MWNTs. We have never observed the monotonic decrease in the conductance reported by Collins et al[67][12], (not even for surfactant coated tubes).

The linear rise in G with increasing V is most likely related to the increase in the density of states with increasing V which also increases linearly with increasing energy[42], as shown in Fig. 9. In fact the DOS of the nanotubes are probed in scanning tunneling spectroscopy. However, for low resistance contacts the increase in conductance is related to the number of accessible channels N, which is the number of 1D subbands that fall within ±1/2$V_{bias}$ of the Fermi level[51]. The conductance G is given by the Landauer equation Eq. 5. Figure 9 gives G(V) assuming the ideal case where T=1 for all channels. In that model, for a 13.6 nm diameter nanotube, we expect that the conductance increase is dG/dV=12 $G_0$/V (Fig. 9). However, the observed increase is much less: dG/dV~0.3-0.5 $G_0$/V. This implies that T~0.02-0.03 for all of the semiconducting subbands while, as shown below T~0.5 for the conducting subbands. The reduced transmission for the semiconducting subbands compared with the conducting subbands are in line with their predicted [47] and observed[48] properties as discussed above.

The strongly reduced transmission of the semiconducting bands reflects the scattering along the tube combined with the contact impedance (possibly due to Schottky barriers). If the former dominates, then the mean free path for the semiconducting subbands is $L_0$= 0.02 L, where L is the nanotube length from contact to contact (from TEM studies L is found to be of the order of 5-10 μm), so that the mean free path $L_0$~100-200 nm. Note that this value is in fact close to the mean free paths found in by Schonenberger et al.[13]. In this case, this indicates the participation of the semiconducting bands to the transport as in fact has been found to be the case in other work[13][16]. Anantram[68] investigated nanotube transport as a function of bias and found reduced transmission coefficients for the semiconducting subbands (crossing bands) compared to the metallic subbands (non-crossing bands), which correspond to the experimental values.

Alternatively, it may be assumed that for high bias tunneling from the contacts to deeper conducting layers occurs so that those layers participate in the transport. This picture is however contradicted by the pattern of destruction at high bias, where a uniform layer is removed from the entire length of the nanotube, which appears to imply that only the top layer participates. Moreover, the next conducting layer is statistically most probably separated by two or more semiconducting layers (i.e. by about 1 nm) which is rather large[40]. Also, the number of the semiconducting spacer layers varies from one MWNT to the next in contrast to the G(V) behavior which we find is quite uniform from one tube to the next. For these reasons we believe that the characteristic linear rise in conductance is due to the participation of the semiconducting subbands of the outer (conducting) layer only, and that these semiconducting subbands have small transmission coefficients (see Eq. 5).

In-situ TEM experiments have shown several examples where a nanotube is contacted on both sides, however applied voltages up to 10 V (i.e. much greater than the band gap) do not produce a measurable current (R>>1MΩ). These are clearly semiconducting nanotubes, however it is curious that potentials as high as these still do not produce a significant current. For example, tunneling into deeper conducting layers or into the states above the gap should contribute to the transport. Since this does not occur implies that the semiconducting tubes are good insulators with high dielectric strengths.

Surfactant coated tubes show very different G(V) behavior (see Fig. 6). In contrast to clean tubes there is no extended linear region and G saturates at V=-1.5 V. For instance a large offset of 0.3 V in the symmetry axis of G(V) is observed in Fig. 6. All these features are in sharp contrast to clean tubes (Fig. 5). The asymmetry may indicate significant doping caused by the surfactant, causing a shift of the charge neutrality point. From this observation we speculate that the ubiquitous doping[13 69] and the water sensitivity[16] observed in processed MWNTs are not an intrinsic nanotube properties but are a directly related to the surfactants that have been applied to the nanotubes[17].

It is interesting to point out that statistically, for 1 in 3 conducting tubes, the second layer is also metallic. It would be expected that these tubes would have remarkably different non-linear properties at higher bias voltages as well as greater low bias conductances (i.e. 2 $G_0$ rather than 1 $G_0$). This is not seen, all clean conducting tubes behave much alike with a nearly perfect linear increase of the conductance and G~1$G_0$. It may well be that those tubes for which the top two layers are metallic are in fact very poor (diffusive) conductors due to interlayer scattering. Scattering of this kind has been described by Roche et al.[70] This implies that those nanotubes that exhibit ~1$G_0$ conductances, the top layer is always metallic and the next layer is always semi-conducting. This immediately explains the great uniformity in properties of all of the conducting MWNTs and their similarity to SWNTs.

In summary it appears that only the conducting subbands of the outer layer participates to the transport. The higher subbands have short mean free paths[48] and/or higher contact resistances which limits their participation to the transport. Among other things, this explains the uniformity in the MWNT transport properties: the number of metallic subbands is the same for all nanotubes. Surfactants cause doping and reduced transmission.

**Comparisons with theory and with other experiments**

The basic electronic structure of SWNTs was theoretically predicted by Mintmire et al.[71] and later experimentally confirmed by Wildoer et al[44] and Odom et al.[43]. The theoretical prediction of ballistic conduction in carbon nanotubes over microns distances by White and Todorov[47] came later and coincided with Frank et al's paper[1]. They pointed out one dimensionality of the electronic structure and the virtual absence of back-scattering for the conducting subbands which, it was speculated, should lead to exceptionally long mean free paths. This theme was later amplified by others and mainly addressed SWNTs (for a recent review, see [72]). MWNTs were treated by Roche[70] and others, who pointed out the importance on interlayer scattering in conductor/conductor double-walled nanotubes and the absence of scattering in conductor/semi-conductor conductor double-walled nanotubes.

Below we discuss a selected set of key experimental papers that directly address the question of ballistic conduction in MWNTs.

A feature of earlier and some later nanotube measurements is that the measured transport properties were diverse and difficult to rationalize: each MWNT appeared to have unique transport properties. For example, four point measurements by Ebbesen et al[5] on several lithographically contacted MWNTs showed a wide variety of properties, with both positive and negative temperature coefficients of the conductivity. Resistivities varied greatly; even apparently negative resistivities were observed, where the voltage measured on the inner two contacts had a polarity which was reversed from that of the outer contacts. The conclusion was drawn that currents in MWNTs follow complex serpentine paths that may even reverse direction. It was later accepted that the problem with these measurements was in the sample preparation. It should be pointed out that the measurements showed signs of poor contacts: the reversed voltage is more aptly explained in terms of a directional mesoscopic contact[51]. However the fact remains that these measurements on lithographically contacted nanotubes yielded unreliable results which, if not explained and corrected, should signal that great caution should be taken in applying similar methods to extract nanotube properties.

Measurements by Langer et al[6] on MWNT bundles showed lnT dependence, which saturates at low temperatures (the conductance increases by about a factor of 2 from 1 K to 80 K). Magnetoresistance measurements showed evidence for universal quantum fluctuations and weak localization. These measurements strongly supported that isolated MWNTs behave as disorder mesoscopic 2-D systems. Weak localization requires that elastic scattering dominates inelastic scattering, and phase coherence lengths greater that the elastic scattering lengths. Hence, these experiments provide evidence for elastic scattering in the tubes.

Measurements by Schonenberger et al[13] on individual MWNTs found closely related results. The nanotubes were purified and ultrasonically dispersed in liquid using surfactants as described in Ref.[17]. The conductance increased by a factor of about 2 when the temperature is increased from 1K to 80 K. Magneto-transport measurements also showed universal quantum fluctuations and weak localization. Observations of Aharonov-Bohm oscillations showed that only the outer layer participates in the transport[49]. Moreover, from tunneling spectroscopy the electronic structure was confirmed to be similar to that of a SWNT however with the expected reduced energy scale[13]. They concluded that the transport in MWNTs is one dimensional, diffusive at room temperature and quasi-ballistic at low temperatures. Temperature independent elastic scattering-lengths $l_e$=90-180 nm were deduced. Furthermore, there was no clear signature for electron-phonon scattering up to T=300 K (see also Hertel[73]), and it was concluded that the conductivity increase with increasing T was not due to density of states (DOS) effects. Note that the DOS increases sharply above the gap, which should cause a very large conductivity increase with increasing temperature (which is not observed). The length dependence of the resistance was estimated to be (by comparing different nanotube samples with different lengths) about 4 kΩ/μm.

More recently, Buitelaar et al.[16] observed quantum dot properties in MWNTs, similar to those observed in SWNTs [8]. It was concluded that the outer layer was disordered with substantial hole doping and that the next layer was metallic to produce the observed properties which were clearly associated with 2 conducting channels from the deeper layer. Coherent transport was assumed (at sub 1 K temperatures) over the entire tube length of 2.3 μm. Substantial hole doping has also been concluded by that group in other work[74] so that up to 10-20 1 D modes of the outer layer participate in the transport, but that charge transport to the contacts is determined by only one mode[69]. The doping has been identified to be related to water[69]. It is also significant that the two point conductances of their MWNTs do not exhibit the increase with increasing voltage [75] that we observe, and that these tubes also exhibit the failure behavior found at high voltage by Collins et al. [12]

Liu, Avouris et al[55] report on the transport properties of two 1% boron doped lithographically contacted MWNTs, which causes a lowering of the Fermi level $\Delta E_F \leq -0.1$ eV. They estimate that 4 and 6 subbands (for the two samples respectively) participate to the transport: The two point 300 K conductivity is found to be G=2.24 $G_0$ and G=2.84 $G_0$. In contrast to others, their samples do not show a decrease but rather a slight linear increase of the relative resistance $R/R_{300K}$ with increasing the temperature from about 100 to 300 K (both in 2 point and in 4 point measurements). However, the increase is extremely small: about 1 $10^{-4}$ /K (a factor of 400 less than for copper). The resistance increase is presented as evidence for metallic conduction. 1D weak localization is concluded from magnetoresistance measurements. The elastic mean free path is found to be $L_{el}$ =220-250 nm which is consistent with scattering only at the contacts. It is estimated that 4-6 channels participate to the transport in these doped nanotubes. The electron-phonon relaxation time at room temperature is estimated to be τ=0.4 ps (which, with a Fermi velocity of $10^8$ cm/sec corresponds to a mean free path of 400 nm). Coherence lengths are found to be temperature dependent and longer than the intercontact distance (250 nm) at low temperatures. One of the conclusions of this paper is that the mean free paths are very long despite the rather heavy boron doping. In many respects this work appears to confirm ballistic conduction (at least on the 400 nm length scale), even in the very unfavorable condition of heavy doping, however the paper actually classifies the nanotubes to be in the diffusive regime. The very weak increase in the resistance is all the more important since it implies that the thermally activated subbands apparently do not significantly contribute to the conductivity with increasing T: it shows that there is no large change in the number of participating layers as the temperature is increased. This appears to be consistent with the relatively small observed increase of the conductivity at high bias voltages mentioned above.

Collins and Avouris et al[67] [12] find complex conduction behavior for lithographically contacted MWNTs. The nanotubes were applied to prepatterned Au electrodes, after dispersing them in dichloroethane, centrifugation and a thermal treatment. The transport properties were interpreted in terms of the interplay of the contributions from multiple semiconducting and metallic layers where up to 8 layers contribute to the transport in the high current (non-linear transport) regime[67]. In later work the authors conclude that many shells participate to the transport even at low bias[12]. They observe that the conductance monotonically decreases with increasing voltage. The in air breakdown occurs at relatively low power (320μW), and proceeds in steps of 12 μA; the tubes ultimately fail in the middle[67]. The two point low bias conductance is 3.7 $G_0$ for a 200 nm long tube.

It is not obvious how to reconcile these measurements with the properties presented by the same group in their earlier work mentioned above[55]. In fact, our own measurements could hardly be more different. We always observe a linear increase in the conductance, never a decrease; we do not observe (low bias) conductances greater than 1 $G_0$ nor do we observe the breakdown in steps. Furthermore, our nanotubes (with contacts) can sustain powers up to about 5 mW, and their breakdown pattern involves only the outer layer(s); failure occurs at the contact and not in the middle of the tube. We must conclude that our nanotubes and those investigated by Collins et al[37] must be essentially different objects and the most significant difference is in the processing (most likely due to oxidation damage caused in the thermal annealing step), since Collins et al used nanotubes produced by us in some of these studies[37].

In measurements that in principle are most closely related to those presented here, Bachtold et al[14] measured the voltage drop along MWNTs and SWNT bundles using scanning electrostatic force microscopy of lithographically contacted nanotubes. From their observations, the voltage drop along current carrying nanotubes was determined, from which the resistance per unit length was deduced. They found that the room temperature resistance of MWNTs is $\rho = 10$ k$\Omega$/$\mu$m, while $\rho < 1.5$ k$\Omega$/$\mu$m was found for the SWNT bundle (although inspection of their data appears to show that a significant voltage drop along at least 50% of the 2$\mu$m long bundle). They concluded that SWNTs are ballistic conductors (from the Landauer equation, assuming that the SWNT bundle contained one conducting nanotube with 2 channels, and that the voltage drop occurred at the contacts) and MWNTs are diffusive conductors. The conclusion was based on the ballistic conduction criterion applied to a (hypothetical) 1$\mu$m long nanotube (Eq. 6).

Ballistic conduction has recently been observed in SWNTs[10][11] from quantum oscillations in a Fabry-Perrot experiment implying long elastic lengths and phase coherence lengths (at least the intercontact spacing, 200 nm). These experiments show that (phase coherent) ballistic conduction at 10 K does in fact occur. Room temperature two-point resistances as low as 7 k$\Omega$ have been measured suggesting low scattering at room temperature as well. It is relevant that the SWNTs in this experiment were produced in situ and not chemically or mechanically treated.

The reasons for these discrepancies between the various nanotube measurements needs be clarified. There is very strong evidence that processing indeed alters the properties[24] in particular of the surface layers[16]. Surfactants are universally used to suspend nanotubes in liquids in order to deposit them on substrates. Surfactants chemically bind to the surfaces and may be very hard to remove; in order to remove them may require a high temperature 'annealing' treatment[21] which can cause further damage them[76]. We have directly demonstrated that surfactants greatly increase the resistivities and affect the doping levels. In fact we find that the resistivities of surfactant treated MWNTs are of the order of magnitude observed by others.[13] Clearly water sensitivity[69] may be explained as a result of the hydrophilic surfactant layer on the nanotubes.

Ultrasound has been found to damage nanotubes[24]. Ultrasonic dispersion of the nanotubes is also universally applied to MWNTs in order suspend them and to separate them from the nanotube fiber bundles.

Thermal treatments are used to open the nanotubes by oxidizing the ends. However very similar treatments are used to burn away graphitic particles and amorphous carbonaceous material and also to anneal the tubes. This purification method clearly can be detrimental to the transport properties and may partly explain the properties observed in Refs.[12, 37]

**Summary and Conclusion**

This comprehensive treatment of the properties of freely suspended unprocessed nanotubes contacted with a liquid metal contact shows that MWNTs are indeed ballistic conductors at room temperature over many microns as originally claimed. Ballistic is meant in the sense that there the momentum scattering lengths are much longer than the nanotube length, hence that the resistance is essentially independent of the length[51][50][14].

The conductance measurements of MWNTs have shown several uniform, robust and reproducible properties: (1). Rounded conductance steps followed by plateaus are always seen. (2) Initial plateau conductances are distributed primarily in the narrow range from $G_{pl} = 0.5$ -1 $G_0$. (3) Initial plateaus significantly greater than 1 $G_0$ are not observed. (4) The great majority of the plateaus are remarkably flat, without small substeps or slopes. (5). Conductances are independent of voltage up to about 0.1 V followed by a linear increase with increasing voltage. (6) Destruction occurs at currents of the order of 1 mA and failure occurs at one of the contacts. (7) The properties of conducting nanotubes are very uniform.

The obvious reason for the uniformity in the properties is given by the theoretical prediction that (for $V_{bias} < E_{gap}$ and $kT < E_{gap}$) only two conducting subbands contribute to the transport for conducting nanotubes. These conditions are amply met for the nanotubes in this study at room temperature and for $V_{bias} < 100$ mV. The linear increase in conductivity at high bias is also clearly explained in terms of participation from higher subbands with reduced transmission. In 2/3 of the cases, the layer below the top layer is semiconducting and hence is not expected participate to the transport, in line with experiments that show that only the top layer participates. Hence, the most straightforward explanation for all these effects is that the two conducting subbands of the outermost conducting layer dominate transport at low bias and at room temperature. As pointed out, our data at low bias strongly disagree with interpretations that attribute the high conductances as due to the participation of many highly resistive conducting subbands. Moreover, high doping levels are not indicated in particular by the very small range in the measured conductance values: doping concentrations are bound to vary and the resistances are expected to diameter dependent. There is no indication for these dependences.

The analysis of the conductance trace shapes shows that the nanotube-metal contact resistances dominate the shape. This contact conductance is rather small: $G^* \sim 50\ G_0/\mu m$ (a contact 100 nm in length has a resistance of about 2 k$\Omega$). The nanotube-metal contact conductances are consistent with other measurements, with recent calculations[64] and with Sharvin's semi-classical expression for contact conductances of small contacts[63].

The nanotube-fiber contact appears to have a transmission coefficient T =0.5, which compares well with the expected limits from T= 0.3 (derived from SWNT junctions) to T=0.7 (the theoretical maximum). Hence, the 'missing quantum of conductance' may be caused by reflections at the nanotube-fiber contact, although we believe that a deeper explanation is not ruled out. Variations in the plateau conductance have been shown here to be caused by variations in that contact. The relatively high frequency of T~0.5 $G_0$ may be due to a scattering at a pseudocontact, i.e. a graphitic flake on the tube which essentially reduces the transmission by a factor of 2 [51].

The slopes of the conductance plateaus are related to the contact resistance and to the resistance per unit length of the tube. The resistance per unit length is found to be $\rho$ <50 $\Omega$/$\mu$m. Combined with the conclusion that two subbands participate in the transport, implies that $l_m$>200 $\mu$m, following the identical reasoning presented by McEuen and co-workers[14]. Hence MWNTs are certainly ballistic conductors at room temperature.

Elastic scattering lengths (from static scattering sites) of the order of hundreds of nm (as estimated for the elastic mean free paths by others[13]), should have produced observable steps in the conductance plateaus which are not seen. Hence either there are no defects on the cleaned tubes or they have a negligible effect on the resistance. Surfactant coated tubes shows resistances of the order of 2 k$\Omega$/$\mu$m. The anomalous G(V) behavior of surfactant coated tubes further indicate doping. Hence surfactant coated tubes are doped with reduced mean free paths compared with clean tubes. These observations are consistent with other measurements which show an increased number of conducting channels, short mean free paths and evidence for doping[69,16].

The conductance versus voltage measurements of freely suspended nanotubes universally show a rise with increasing voltage, even for surfactant coated tubes. For clean tubes the slope is tube dependent. The increase is clearly explained in terms of the participation of the semiconducting subbands at high bias however with reduced transmission coefficients.

The nanotubes can sustain high currents (order of mA). In–situ TEM experiments verified the conductivity properties of nanotubes. They further show that dissipation occurs at defects and contaminants and failure occurs at the contacts. At high current densities, the outer layer is destroyed showing that only the outer layer conducts as was concluded earlier[1,2,13,49].

In contrast, the properties of processed, lithographically contacted are very different and vary from one experiment to the next. Due to this variety, a uniformly applicable summary of properties cannot be given, and the following properties are representative. (1) The nanotubes are diffusive conductors with low temperature mean free paths of the order of fraction of a micron. (2) The tubes are doped and the transport involves multiple (>2) channels[13,69]. (3) In some cases the transport is complex involving many layers[12], in other cases only the top two layers contribute of which the top layer is doped and the deeper layer shows ballistic properties[16]. (4) The conductance decreases with increasing voltage and the nanotubes fail due to thermal heating at relatively low currents[38].

This comparison clearly demonstrates that the lithographically contacted processed nanotubes are not the same objects as the unprocessed freely suspended nanotubes that we have investigated. We believe that processing damages in particular the outer layers of the nanotubes which are the most important ones for electrical transport. Our experiments abundantly demonstrate the excellent and unique ballistic transport properties of the multiwalled nanotubes, which still are unrivaled in any other system. More importantly, we have demonstrated that these unique quantum properties persist under ambient conditions.

The measurements by Frank et al[1] were the first to demonstrate not only ballistic conductance in virgin carbon nanotubes under ambient conditions but also their 1D properties, their high current carrying abilities, and the fact that only the outer layer conducts. These properties were found at a time when there was no indication for any of them either from theory or from other experiments. These properties are in line with those expected theoretically for defect free nanotubes, and also in line with more recently found properties of SWNTs.

Acknowledgements: C.B. is grateful to the French CNRS & to NATO for financial support, and thanks Joe Gezo for his help with the electron microscopy experiments. Financial support for this project was provided by the Army Research Office: DAAG 55-97-0133; and by the National Science Foundation: DMR 9971412 and is gratefully acknowledged.

**Appendix**

A semi-classical model for a multi-element conducting system is developed here, following a similar treatment presented in earlier works[60, 61] and a related treatment by Datta.[51] The essential feature of this highly simplified model of transport in a mesoscopic wire connected to reservoirs is the following (see Fig. 10). Each element is represented by its number of channels (i.e. conducting 1D subbands that intercept the Fermi-level). Electronic scattering at an interface between two elements is isotropic and elastic: electrons scatter with equal probability into all the accessible channels while conserving energy (the isotropy assumption, cf Beenakker[62]). The transmission probability of the system is found by considering all trajectories, but ignoring interference effects.

For example, consider a wire with $n$ channels is connected to a contact with $m$ channels and one with $p$ channels. These contacts are in turn connected to reservoirs with an infinite number of channels. As shown earlier, this model accurately predicts conduction histograms in break junctions (nanowires). In short, at the interface between the contact and the wire electrons are scattered elastically and isotropically into all possible channels (the isotropy condition[62]). Hence at the interface of the wire with the second contact, an electron in one of the n channels of the wire scatters with equal probability into the p channels of the contact as well as back into the n channels of the wire. Consequently, the transmission probability of that electron from the wire into the contact is $p/(n+p)$; *t*he probability that it reflects is $n/(n+p)$. Hence, considering multiple scattering, summing the resulting infinite series, and using the Landauer equation for conduction [50, 51], yields a remarkable simple expression for the conductance of this wire

$$G = G_0/(1/n + 1/m + 1/p) \quad (A1)$$

This result can be generalized to any 1D system with an arbitrary number of contacts and scattering centers[60, 61]:

$$G = G_0 \left( \Sigma(n_i^{-1}) \right)^{-1} \quad (A2)$$

This provides a simple way to estimate the transport through a system of scatterers and contacts[52]. In fact, as shown in[60, 61], if one assumes that n, m, and p take any value from 1-20, say, and one produces a histogram of the conductances, reproduces very nicely the conductance histogram of breaking nanowire contacts, without resorting to an arbitrary series resistor (see Fig. 11). As discussed in Refs.[60, 61], this provides a natural explanation for the 'serial resistance' in these break junctions: these wires consist of several connected segments with scattering at the junctions.

For a carbon nanotube connected to two non-reflecting contacts, m=2 and both n and p are very large, hence $G=2G_0$.

A nanotube of length L with a mean free path of $L_0$, has $L/L_0$ scattering centers (which in this model are assumed to scatter electrons elastically and isotropically), and the conductance will consequently be [51, 14]

$$G_{tube} = 2G_0 /( L/L_0 + 1) \quad (A3)$$

The nanotubes are contacted on one side to the liquid metal contact, which is represented by $n_1$ channels; $n_1$ is estimated from the Sharvin equation[63] as the area of the contact divided by the Fermi wavelength squared. In any case, it is proportional to the contact area and hence to the contact length x, this contact is represented by its number of channels: $n_{C1} = C_1 x$. For the moment, we consider that the contact to the fiber is ideal with a transmission probability of 1.

Consequently the resistance of a system consisting of a nanotube with two channels of length L, connected to a metal contact (length x) and a fiber contact represented by $n_{C2}$ is

$$R(x) = (2G_0)^{-1} ((C_1 x)^{-1} + ((L-x)/L_0 + 1)) \quad (A4)$$

This result converges to the expected values at the extremes as is easily verified. Moreover, the functional dependence on x is as expected in the classical limits and justifies the shape analysis presented in this work. It is interesting to note that the experimental value of $C_1$ is in fact of the same order of magnitude as predicted from the Sharvin equation.

Next, we address the contact of the nanotube with the fiber bundle. The contact of the nanotube to the fiber bundle is a series of contacts to other nanotubes, as schematically shown in Fig. 10. At a contact point an electron can scatter back, continue forward or transfer to the crossing nanotube. For example, in keeping with the previous discussion, we assume that the each of the three possible scattering directions have equal probability. By summing the resulting infinite series of possible paths one finds that the transmission coefficient for the nanotube-fiber contact

$$T_{C2} = 2/(1+\sqrt{5}) = 0.69 \quad (A5)$$

This should be considered to be the maximum possible transmission coefficient for electrons entering into the fiber from the nanotube.

Recently crossed single walled nanotube junctions have been studied explicitly and high transconductances, with $T_{12} = 0.06$ for tunnel from one metallic tube to the other. The general expression for the contact conductance $T_{C2}$ (after summing the series) is

$$T_{C2} = 2/(1+\sqrt{2/T_{12}-1}); \quad (A6)$$

Hence $T_{C2}=0.3$ for $T_{12}=0.06$. Note that our measurements imply that $T_{C2}=0.5$ which would require $T_{12}=0.2$. This value is below the maximum estimate ($T_{12}=1/3$) and above the largest value found for crossed SWNTs. In any case, it is reasonable to expect that $T_{12}$ for MWNTs is somewhat greater than for SWNTs.

† Permanent address: GDPC, UMR 5581,CC026-Université Montpellier II, 34095 Montpellier Cedex 5,France;
†† Permanent address: CNRS-LEPES, BP 166, 38042 Grenoble Cedex9, France
* To whom correspondence should be sent

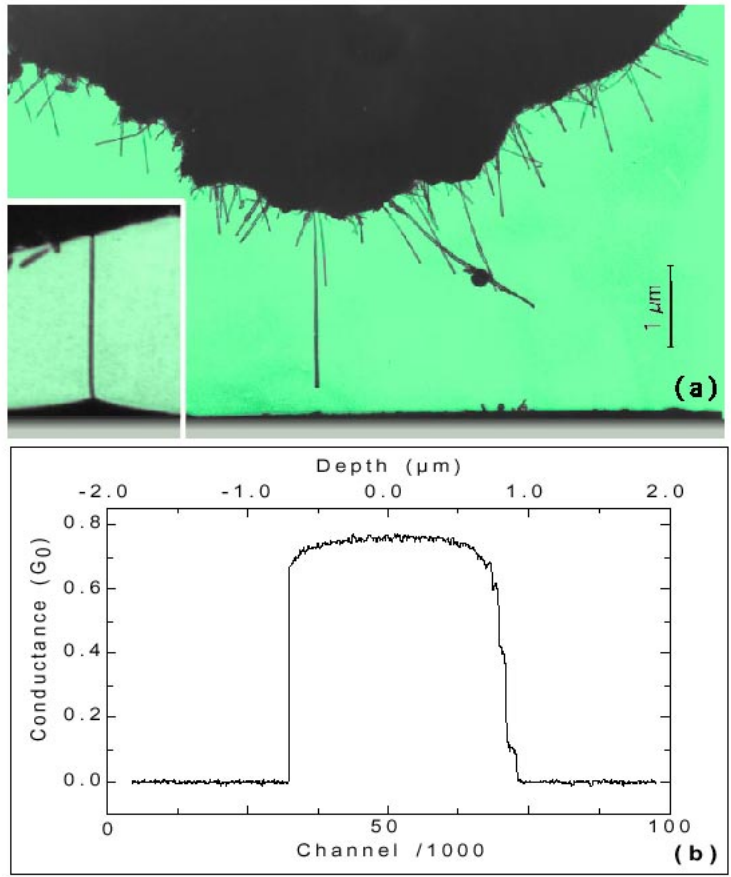

Figure. 1 : TEM image of a multiwalled carbon nanotube fiber tip opposing a mercury surface and the dipping process. **(a)** The nanotubes protrude from the fiber that is composed of densely packed carbon nanotubes and other graphitic nanostructures. The transport measurements are made by lowering the tip into the liquid metal and measuring the conductance as a function of the position. **Inset** : Example of cone shaped meniscus attached to the tip of the nanotube which occurs when the nanotube is pulled out of the (non-wetting) liquid just before contact is broken. **(b)** A full cycle conductance trace (conductance G=I/V versus position) where the fiber is first lowered to the Hg and subsequently withdrawn (see upper axis). Note the asymmetry with respect to the turning point due to the non-wetting adhesive effects.

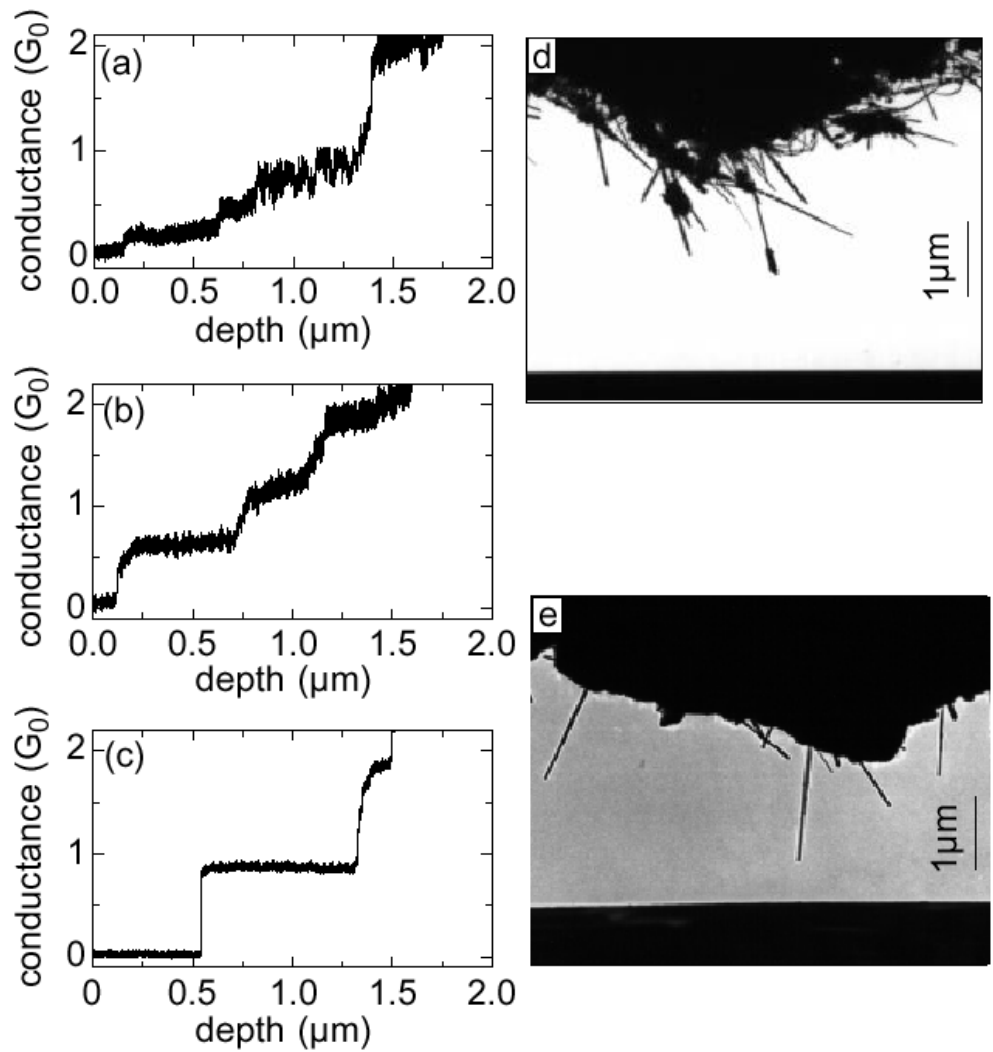

Figure 2. Cleaning of nanotubes and evolution of nanotube fiber properties by repeated dipping in Hg. (**a**) Conduction trace of the virgin fiber: steps are barely discernable; (**b**) Steps develop after a few hundred cycles but they still exhibit relatively large slopes and jumps (**c**) After several thousand cycles, the steps are well developed and the pattern is stabile. The first step evolves from the shoulder seen in **a** (step: 0.2 $G_0$, slope: 36 k$\Omega$/$\mu$m) to a rounded step in **b** (step: 0.62 $G_0$, slope: 4k$\Omega$/$\mu$m), to the well defined step with a flat plateau (**c**) The second step is due to another tube and evolves analogously. (**d**) TEM Micrograph of a virgin fiber tip opposing Hg surface; note the contaminating graphitic particles and the loose structure of the tip. (**e**) TEM Micrograph of a fiber tip that has previously been repeatedly dipped in Hg; the nanotubes are straight and free of particles and the fiber is compacted.

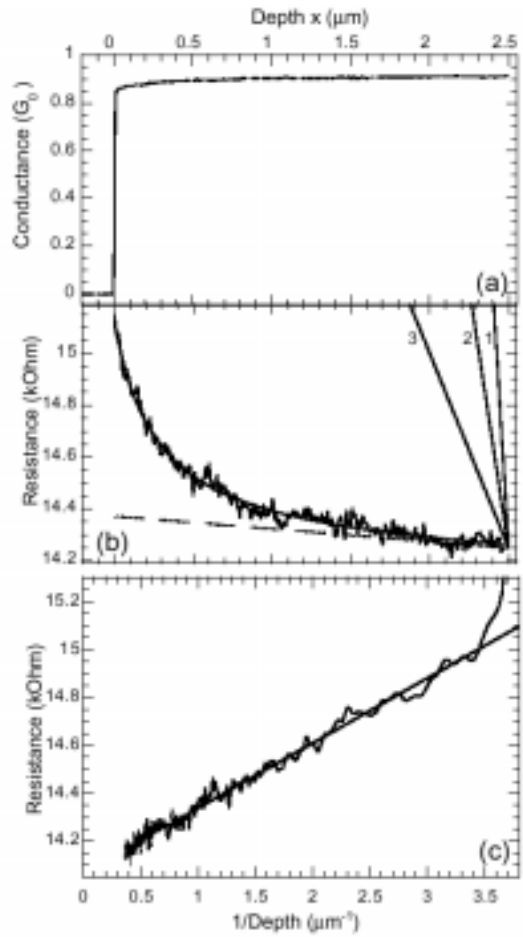

Figure. 3 A representative conductance trace (one of 50 of this nanotube) as a function of the distance x between tip of the nanotube and the Hg surface (i.e. the depth). **(a)** The conductance G(x) in units of the conductance quantum, showing the initial conductance jump at x=0 to 0.85 $G_0$, followed by a rounded step, of which the slope gradually decreases to 0 with increasing x. **(b)** The resistance R(x)=1/G(x). Note that the slope gradually decreases to 0. Dashed line corresponds to the slope at x=2.5 µm, which corresponds to the upper limit of the tube resistance: $\rho < 48$ Ω/µm; line (1) corresponds to $\rho$ =10 kΩ/µm found in Ref. [14] for MWNTs; line (2) $\rho$ =4 kΩ/µm as in Ref. [13], and line (3) $\rho$ =1.5 kΩ/µm found for a SWNT bundle which was characterized as a ballistic conductor (Ref. [14]) **(c)** Nanotube resistance plotted as a function of 1/x, revealing a straight line: R(1/x)=14.1+0.271/x kΩ. This demonstrates that the contact resistance indeed determines the shape of the conductance trace.

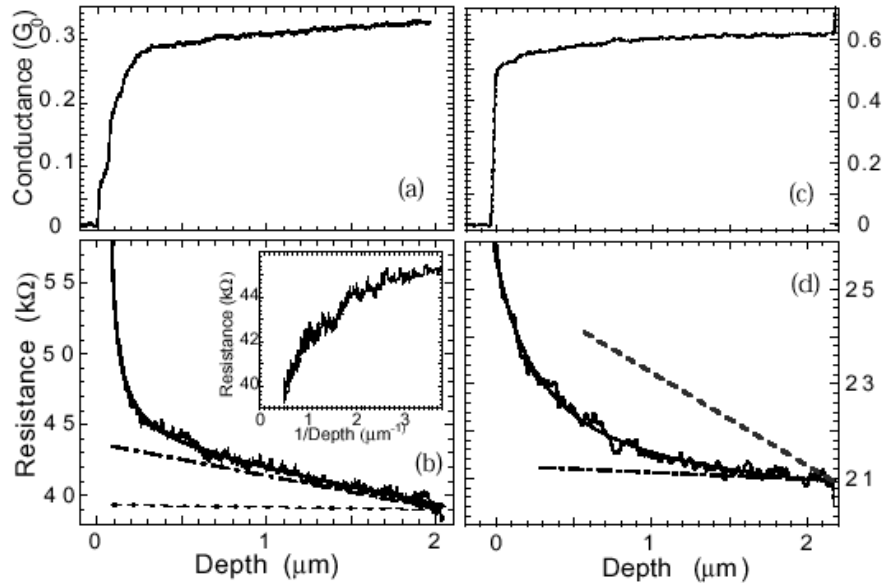

Figure. 4. Effect of a surfactant on G(x). (**a**) G(x) for a sodium dodecyl sulfate coated nanotube; in contrast to clean tubes the conductance continues to increase with increasing x. (**b**) solid line R(x)=1/G(x); **full line**: fit to the semi-classical model (Eq. 3). Note that R(x) asymptotically approaches the slope –dR/dx = 2.3 kΩ/μm (dot-dashed line), from which about 0.1 kΩ/μm is due to the metal contact resistance of 500 Ωμm. The –dR/dx slope is more than an order of magnitude greater than the slope typically found for clean tubes of similar length i.e. –dR/dx = 0.2 kΩ/μm (dashed line).Inset : Resistance R as a function of 1/x, showing that contrary to a clean tube the shape of the conductance trace is not only determined by the contact conductance. (**c**) Example of G(x) for a clean nanotube of similar length (2 μm), with a low plateau conductance $G_{pl}$=0.64 $G_0$ (**d**) R(x) asymptotically approaches the plateau resistance $R_{pl}$= 20.3 kΩ ; from the slope at x= 2μm, -dR/dx =260 Ω/μm (dot-dashed line) which is an upper limit to the resistance of the tube. Also shown is the slope -dR/dx= 2.3 kΩ/μm (dashed line) corresponding to the surfactant coated tube in (**b**).

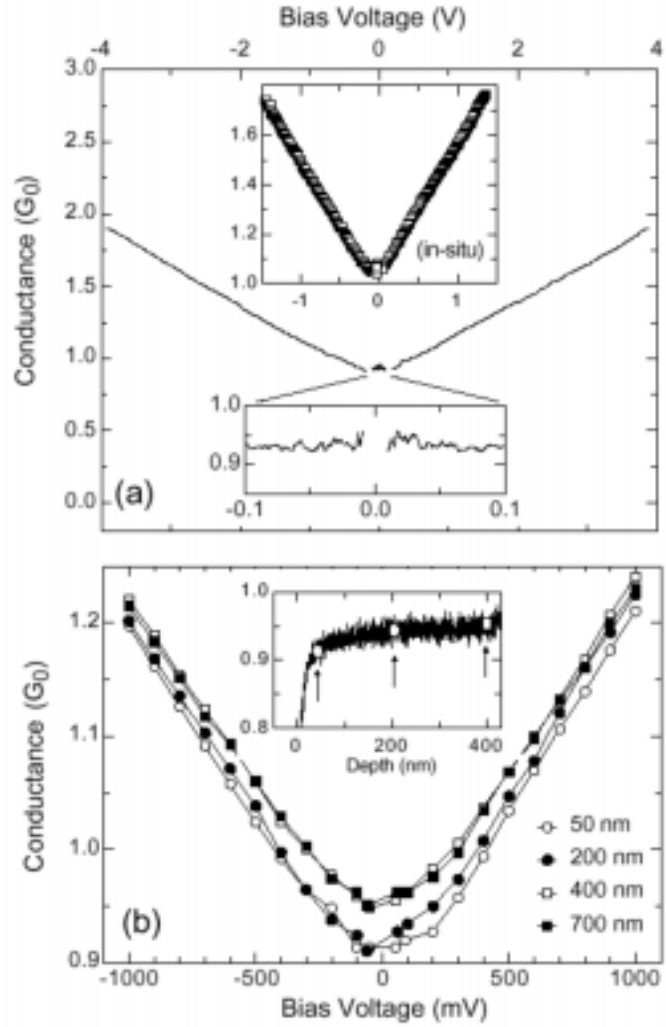

Figure 5 Conductance versus voltage. **(a)** G(V) for a clean nanotube in air from V= – 4.0 to +4.0 V and from V=–1.3 to +1.3 V in the TEM (inset). Note the striking symmetry and the essentially perfect linearity of G(V). This is a robust property of the nanotubes studied in these experiments. Note that there is no evidence for saturation and certainly not for a decrease in conductivity with increasing voltage. The current at V=4 V corresponds to I=620 µA. **Open circles** : in situ measurements of G(V) of a nanotube contacted in the TEM. **(b)** G(V) of a nanotube for various positions x into the Hg as indicated on the G(x) trace in the inset.

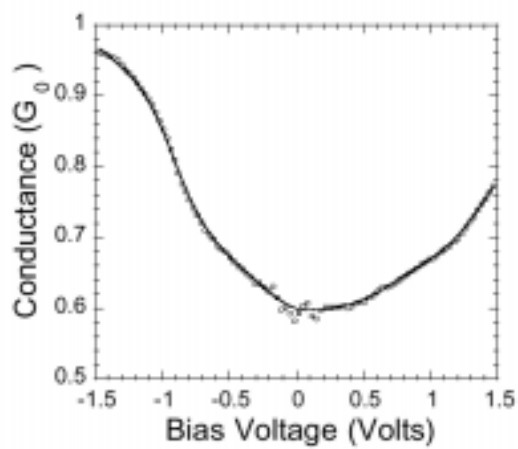

Figure 6. Effect of a surfactant on G(V) : an example of G(V) for a surfactant coated nanotube. Note the differences with clean tubes, in particular the non-linearities and the asymmetry with respect to V=0. The conductance saturates for V=-1.5 V. This behavior is reproducible of this nanotube, however the shift and the asymmetry is sample dependent.

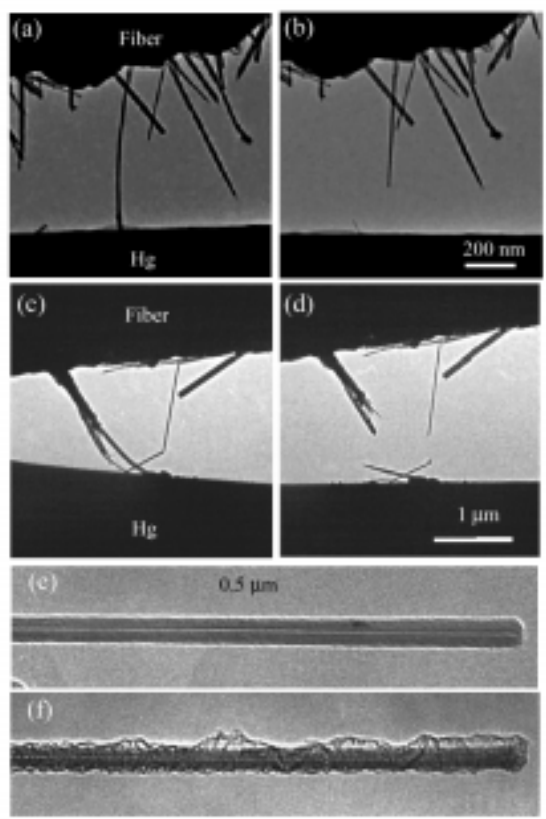

Figure 7. Before and after in situ TEM images of contacted nanotubes and their failure at high currents. **(a-b)** Typical failure of a clean nanotube. The failure occurred at the contact with the Hg after applying 4 V leaving a short (~20 nm long) stem at the original contact point. Before the failure the measured resistance was 12.7± 0.2 k$\Omega$. **(c-d)** One kinked and two contaminated nanotubes, showing that the failure occurred at the defects. **(e-f)** High resolution images of the failure of a clean nanotube showing that only the outer layer is affected, which corresponds with the current flow pattern in these tubes.

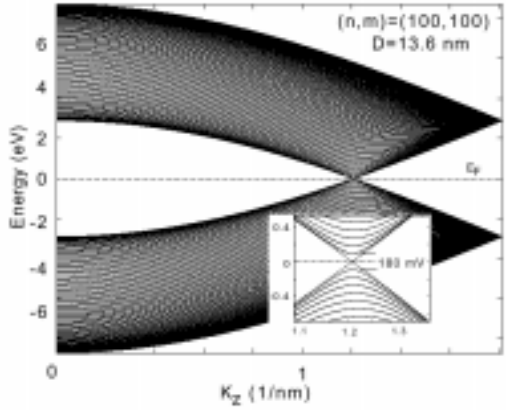

Figure 8. The 1D electronic bands of a (n,m)=(100,100) carbon nanotube calculated in the tight binding model with $\gamma_0$=2.9 eV and a=0.142 nm, where the energy of the subbands are plotted versus $K_z$ (the wave number along the tube). This D=13.6 nm diameter nanotube is in the range of diameters typical of the nanotubes studied here (i.e. 5nm<D<20nm). The electronic transport in this metallic nanotube is due to the two subbands that cross the Fermi level (see inset). Above and below the Fermi level are two sets of semiconducting subbands. The gap between these is $E_{gap}$=6 $\gamma_0$a/D=180 mV (~7 kT at room temperature, note that for semiconducting tubes with the same diameter, the gap is a factor of 3 smaller). The transport properties of the conducting subbands are unique and characterized by very low back scattering compared with the semiconducting bands.

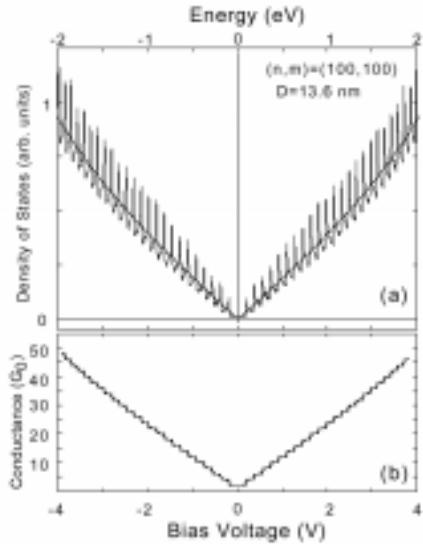

Figure 9. The density of states versus energy of the nanotube in Fig. 8.(**a**) The typical van Hove singularities, which occur when the energy coincides with the bottom of the subbands (Fig. 8), produce a set of approximately equally spaced spikes. Superimposed is also the DOS after gaussing smoothing with $\Delta E$=25 mV to simulate effect of room temperature. This results in a nearly linear dependence of the DOS with energy. For $|E|<E_{gap}/2$=90 mV the DOS is essentially constant. (b) The predicted conductance G versus bias voltage for this nanotube from the Landauer equation, assuming unit transmission for all channels, which states that when the bias voltage increases above the bottom of a subband, then that subband contributes $G_0$ to the conduction, which gives G(V) its staircase appearance. Due to the symmetry above and below $E_F$, contributions from subbands below $E_F$ and above $E_F$ coincide so that the conductance increases in steps of 2 $G_0$. Thermal smearing at T=300 K blunt the steps to provide an essentially perfectly linear rise in the conductivity with increasing bias voltage. The linear increase in the DOS is common to all metallic nanotubes independent of helicity up to about $V_{bias}$= 6V.

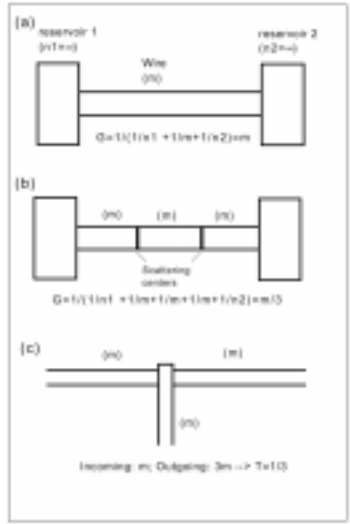

Figure 10. Isotropic scattering of transport through a mesoscopic wire. The transmission at an interface is given by the number $m/(n+m)$ where $m$ is the number of channels in the forward and $n$ in the backward direction. (a) A reservoir is assumed to have an infinite number of channels hence the conductance of the wire is $G=m$. (b) Scattering in the wire divides the wire into connected segments as shown. By summing all trajectories, it can be shown in general that $G=1/(1/n+1/m+...+1/p)$, where n, m, p are the number of channels in each segment. Hence for the two scatterers, $G=m/3$. (c) Demonstration that transmission through a ballistic wire in contact with an other one is $T=1/3$.

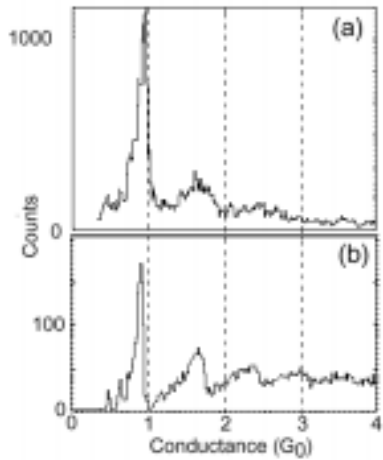

Figure 11 Example of how the semi-classical model accurately describes the well-known features of breaking nanowires. (a) Histogram of the conductance plateaus obtained from several thousand breaking gold nanowires; (b) Histogram of $G=1/(1/n+1/m+1/p)$ for all values of $n, m, p$ from 1 to 20. The model accurately describes the position of the conductance peaks and the general shape of the histogram without requiring the arbitrary series resistor shift that is commonly used to line the peaks up with the expected quantized values.